\documentclass[]{spie}  
\usepackage[dvips]{graphicx}

\title{Sacrificial charge and the spectral resolution performance \\ of the Chandra Advanced CCD Imaging Spectrometer}

\author{Catherine E. Grant, Gregory Y. Prigozhin, Beverly LaMarr and Mark W. Bautz
\skiplinehalf
Center for Space Research,  Massachusetts Institute of Technology, \\ Cambridge, Massachusetts 02139
}

\authorinfo{Further author information: (Send correspondence to C.E.G.) \\ C.E.G.: E-mail: cgrant@space.mit.edu}

\begin{document} 
  \maketitle 

\begin{abstract}
Soon after launch, the Advanced CCD Imaging Spectrometer (ACIS), one of
the focal plane instruments on the Chandra X-ray Observatory, suffered
radiation damage from exposure to soft protons during passages through the
Earth's radiation belts. The ACIS team is continuing to study the properties of
the damage with an emphasis on developing techniques to mitigate charge
transfer inefficiency (CTI) and spectral resolution degradation. A post-facto
CTI corrector has been developed which can effectively recover much of the
lost resolution. Any further
improvements in performance will require knowledge of the location and
amount of sacrificial charge - charge deposited along the readout path of an
event which fills electron traps and changes CTI. We report on efforts by the
ACIS Instrument team to characterize which charge traps cause performance
degradation and the properties of the sacrificial charge seen on-orbit. We also
report on attempts to correct X-ray pulseheights for the presence of sacrificial
charge.

\end{abstract}


\section{INTRODUCTION}
\label{sect:intro}

The Chandra X-ray Observatory, the third of NASA's great observatories in space, was launched just past midnight on July 23, 1999, aboard the space shuttle {\it Columbia}\cite{cha2}.  After a series of orbital maneuvers Chandra reached its final, highly elliptical, orbit.  Chandra's orbit, with a perigee of 10,000~km, an apogee of 140,000~km and an initial inclination of 28.5$^\circ$, transits a wide range of particle environments, from the radiation belts at closest approach through the magnetosphere and magnetopause and past the bow shock into the solar wind.

The Advanced CCD Imaging Spectrometer (ACIS), one of two focal plane science instruments on Chandra, utilizes charge-coupled devices (CCDs) of two types, front- and back-illuminated (FI and BI).  Soon after launch it was discovered that the FI CCDs had suffered radiation damage from exposure to soft protons scattered off the Observatory's grazing-incidence optics during passages through the Earth's radiation belts\cite{gyp00}.  Since mid-September 1999, ACIS has been protected during radiation belt passages and there is an ongoing effort to prevent further damage and to develop hardware and software strategies to mitigate the effects of charge transfer inefficiency on data analysis.  One such strategy, post-facto correction of event pulseheights based on knowledge of the charge history along the transfer direction, is described here.

This paper begins by describing the characteristics of ACIS radiation damage in Section~\ref{sect:raddamage} and of sacrificial charge in Section~\ref{sect:saccharge}.  Section~\ref{sect:data} defines the calibration data used to test correction algorithms.  The basic CTI correction algorithm and the additional sacrificial charge correction algorithm are outlined in Sections~\ref{sect:cticorr} and \ref{sect:saccorr}.  The spectral resolution performance of the correction algorithms are compared in Section~\ref{sect:perf} and further discussed in Section~\ref{sect:disc}.

\section{CHARACTERISTICS OF ACIS RADIATION DAMAGE} 
\label{sect:raddamage}

A symptom of radiation damage in CCDs is an increase in the number of charge traps.  When charge is transfered across the CCD to the readout, some portion can be captured by the traps and re-emitted later.  If the original charge packet has been transfered away before the traps re-emit, the captured charge is ``lost'' to the charge packet.  This process is quantified as charge transfer inefficiency (CTI), the fractional charge loss per pixel.  The front-illuminated ACIS CCDs suffer from radiation damage from low energy protons.  The framestore covers stopped this radiation, so damage was limited to the imaging area of these frame-transfer CCDs.  The charge loss occurs as photons are transfered from the image array to the framestore along the CHIPY direction.    Both the pulseheight and the spectral resolution become position dependent with pulseheight and resolution decreasing with increasing distance from the framestore.  The parallel CTI at 5.9~keV of the ACIS FI CCDs varies across the focal plane from $1 - 2 \times 10^{-4}$ at a temperature of --120$^\circ$~C.  Four species of charge trap have been identified on the ACIS FI CCDs.  The estimated fractional abundances and emission time constants measured at --120$^\circ$~C are of order 4\% with 60~$\mu$sec, 46\% with 400~$\mu$sec, 26\% with 2~msec, and 24\% with $>$~3~sec\cite{bautz}. 

\section{SACRIFICIAL CHARGE} 
\label{sect:saccharge}

Measured CTI is a function of fluence or, more specifically, the amount of charge deposited on the CCD.  As the fluence increases, traps filled by one charge packet may remain filled as a second charge packet is transferred through the pixel.  The second charge packet sees fewer unoccupied traps as a result of the previous ``sacrificial charge'' and loses less charge then it would have otherwise\cite{gendreau}.

Since each X-ray event sees traps with different charge histories, the measured pulseheight for each event can be substantially higher or lower than the average.  The importance of sacrificial charge to the overall performance is determined by the emission time constants of the charge traps and the frequency of sacrificial charge deposition.  During a typical calibration source run on ACIS, an average of 2\% of the pixels have more than 40~ADU of charge - much of that due to cosmic rays.  On average an X-ray event will encounter sacrificial charge of at least this magnitude every 60~rows as it is transfered to the framestore.  One of the ACIS traps has a time constant similar to the average time between sacrificial charge events (2~msec time constant / 40~$\mu$sec pixel to pixel transfer time = 50~rows), so the traps encountered by each event can be in a wide range of occupancy states, which is a source of spectral resolution degradation.

\begin{figure}
\vspace{3.3in}
\includegraphics{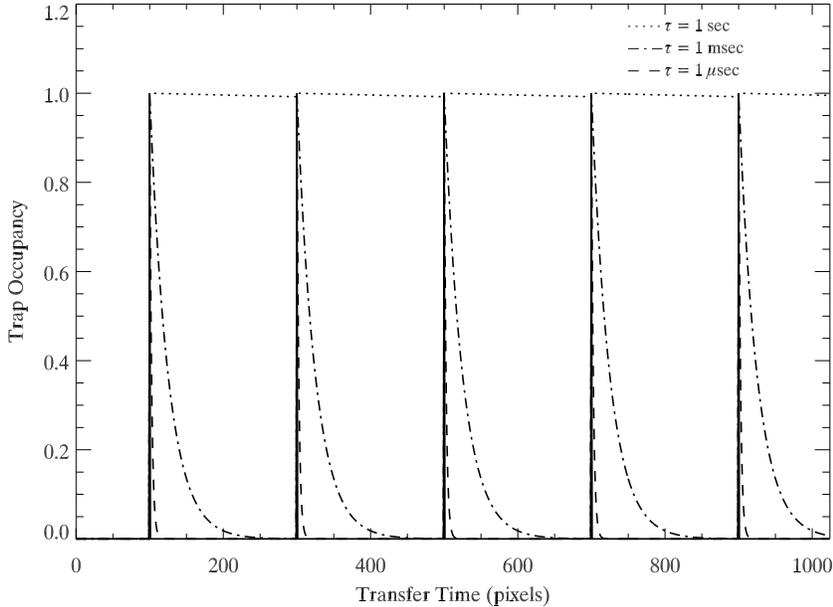}
\caption{An illustration of how emission time constants and sacrificial charge frequency interact.  The idealized input charge distribution fills the charge traps every 200 transfer pixels.  The charge is drained away exponentially with a time constant of 100 $\mu$sec (dashed), 1 msec (dash-dot) or 1 sec (dotted).}
\label{fig:occ}
\end{figure}

Figure~\ref{fig:occ} illustrates how emission time constants and sacrificial charge frequency interact.  The idealized input charge distribution fills the charge traps every 200~transfer pixels.  The charge is drained away exponentially with a time constant of 100~$\mu$sec, 1~msec or 1~sec.  The millisecond trap spends much of the time in intermediate states, while the shortest and longest traps are either empty or full.  The wide variation in millisecond trap occupancy introduces an additional broadening of the spectral resolution.

\section{DATA} 
\label{sect:data}

Because most of the sacrificial charge deposited on the CCD is from cosmic rays that are rejected by on-board event processing, standard ACIS event lists are insufficient for this study and raw data are required.  From May~2001 through March~2002, ACIS has periodically been taking External Calibration Source data in a special telemetry format that packs raw data such that many consecutive data frames can be telemetered.  Primary spectral features of the calibration source are Al-K at 1.5~keV, Ti-K$\alpha$ at 4.5~keV and Mn-K$\alpha$ at 5.9~keV.  Only a 16-column wide strip of ACIS-I3 node 3 is telemetered.  The data used in this study represent approximately 200~ksec taken over 10~months immediately before and after radiation belt passages.  The focal plane temperature was --120$^\circ$~C.  The raw frames were processed to select events, much like on-board, and to record the distance and pulseheight to each of the precursor sacrificial charges for each event.  Only precursors that have more charge than any preceding precursors are included.

\section{CTI CORRECTION} 
\label{sect:cticorr}

Since the charge loss process can be modeled, it is possible to apply a post-facto correction algorithm to replace the lost charge and recover some of the detector's spectral resolution.  Ref.~\citenum{Townsley00} and \citenum{Townsley02} describe a CTI correction algorithm in which a position and pulseheight dependent correction is made to each pixel in the event island (3~x~3 pixels).  The model also accounts for sacrificial charge shielding and charge trailing within the event itself.  A similar algorithm which incorporates this self-shielding model is being implemented by the Chandra X-ray Center as part of \verb+acis_process_events+.  Charge loss in this algorithm is the product of the density of charge traps multiplied by the volume of the charge packet, parameterized as a power law function of energy.  The total charge loss for an isolated pixel is then
$$\delta Q (x,y,Q) = \overline{n}(x,y) \, y \, kQ^\alpha , $$
where $\overline{n}$ is the average trap density of pixels $(x,1)$ through $(x,y)$, $k$ is a normalization constant, $Q$ is the charge in the pixel, and $\alpha$ is the power law index.  Some fraction, $f$, of the charge loss is trailed in to the following pixel.  

CTI correction removes the position dependence of pulseheight, provides substantial improvement in spectral resolution and some small improvement in the detection efficiency.  The improvement in spectral resolution is the result of correcting each pixel in the event island separately and accounting for the spatial variation in the trap density.  The improvement in quantum efficiency is due to the self-shielding model which restores trailing charge into its original pixel thus changing the event grade.  While a CTI corrector can be calibrated to match charge loss induced by differing global levels of sacrificial charge, the spectral resolution degradation caused by sacrificial charge remains.

\section{SACRIFICIAL CHARGE CORRECTION}
\label{sect:saccorr}

We are attempting to develop algorithms to correct X-ray event pulseheights based on their sacrificial charge history.  For each X-ray event $X$, we define $Z_X$, which represents the sacrificial charge history of the event as an equivalent difference in transfer distance induced by the presence of sacrificial charge.
$$Z_X = \sum_{i=1}^{N} \left[ \left(\frac{p_i}{p_X}\right)^\alpha - \left(\frac{p_{i-1}}{p_X}\right)^\alpha \right] (y_X - d_i) \, e^{-d_i/\tau},$$
\vspace*{-0.5cm}
\begin{tabbing}
where \\
aaaaaaaa\=aaa\=\kill \\
$(p_X, y_X)$ \>are the pulseheight and CHIPY of the X-ray event ($p_0 \equiv 0$), \\
$(p_i,d_i)$ \>are the pulseheight and distance to sacrificial charge $i$ (if $p_i > p_X$ then $p_i = p_X$), \\
$\alpha$ \>is a power law index such that $p^\alpha$ is proportional to the volume occupied by charge $p$ ($\alpha \sim 0.5$), and\\
$\tau$ \>is the time constant of the charge traps expressed in rows (1 row = 40 $\mu$s, $\tau \sim 3$ msec = 75 rows). \\
\end{tabbing}
\vspace*{-0.4cm}

\begin{figure}
\vspace{3.8in}
\includegraphics{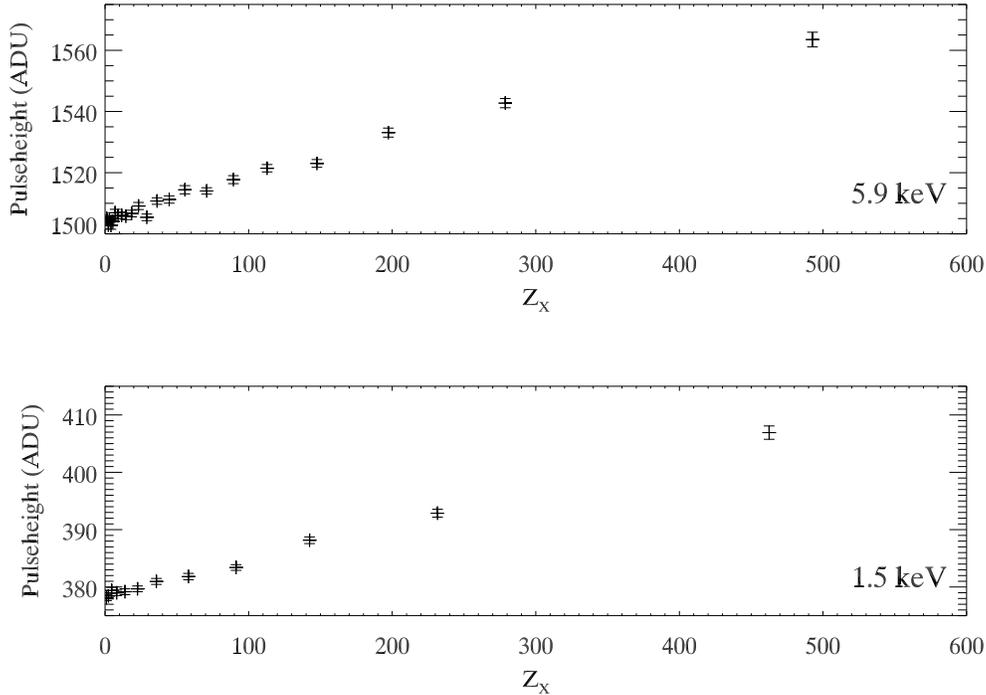}
\caption{Dependence of X-ray pulseheight on $Z_X$, as defined in section~\ref{sect:saccorr}, at 5.9 and 1.5~keV.  Each data point represents a Gaussian fit to 600 X-ray events adaptively binned in $Z_X$.}
\label{fig:phz}
\end{figure}

Events with large values of $Z_X$ are most influenced by sacrificial charge.  Figure~\ref{fig:phz} demonstrates the dependence of pulseheight on $Z_X$ at 5.9~keV and 1.5~keV.  The X-ray pulseheight is nearly a linear function of $Z_X$ as expected.  The modeled charge loss in the correction algorithm is then modified to include the implied change in transfer distance induced by the sacrificial charge as
$$\delta Q (x,y,Q,Z_X) = \overline{n}(x,y) \, (y-Z_X) \, kQ^\alpha . $$
This modification should remove the dependence of pulseheight on sacrificial charge and improve the spectral resolution performance.

\section{PERFORMANCE COMPARISON}
\label{sect:perf}

Our primary figure of merit for judging any CTI mitigation approach is the full width at half maximum (FWHM) of spectral features at the top of the CCD (far from the readout).  On the ACIS-I3 CCD this region corresponds to the ACIS I-array aim point at row 964.  Only the ASCA grade 0,2,3,4, and 6 events are included.  The ACIS External Calibration Source includes many spectral features; the three strongest are Al-K, Ti-K$\alpha$ and Mn-K$\alpha$ at 1.5, 4.5 and 5.9~keV.

\begin{figure}
\vspace{7.9in}
\includegraphics{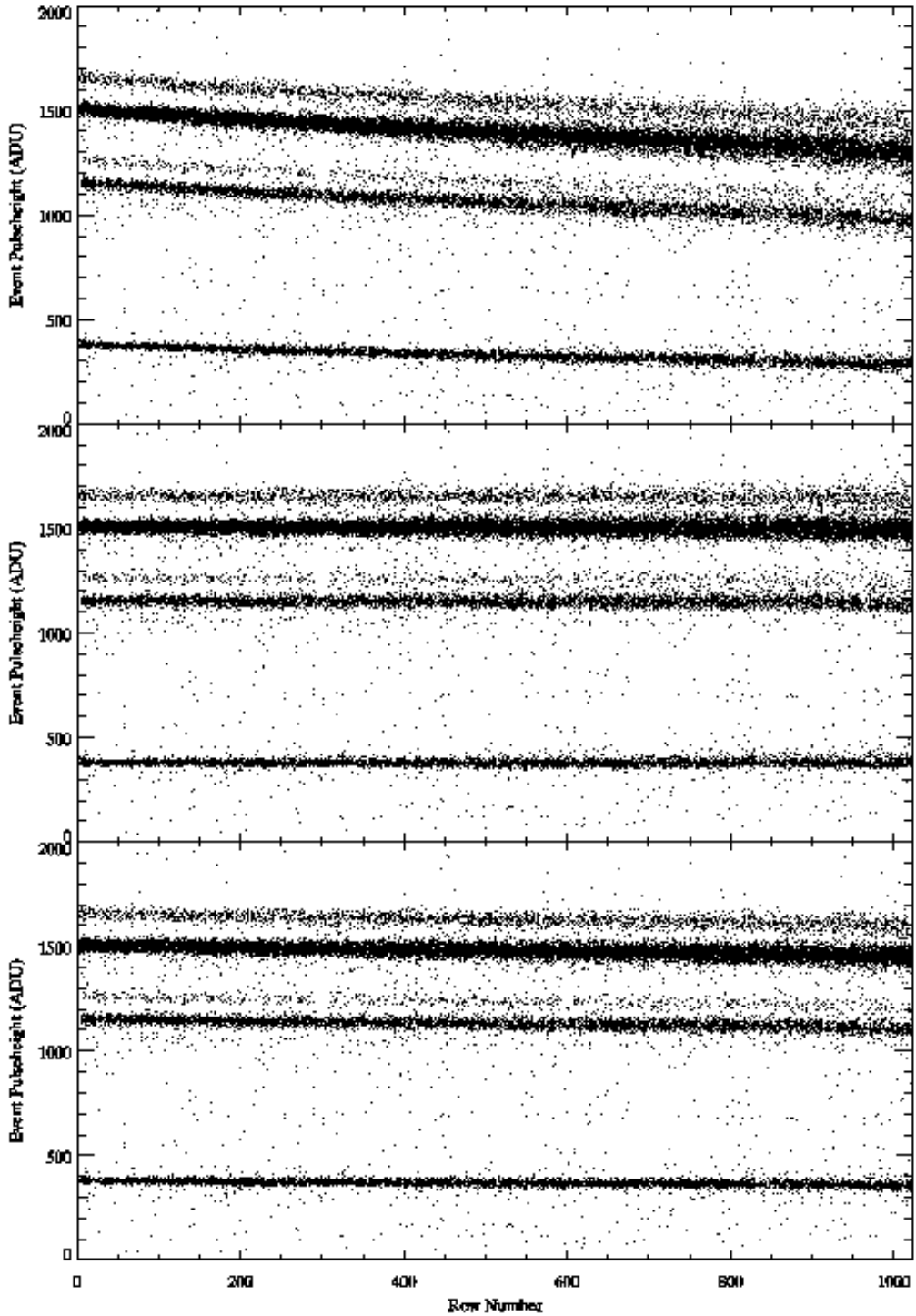}
\caption{This figure shows the pulseheight of each X-ray event versus its row number for data that is uncorrected (top), has been corrected for CTI (middle) and has been corrected for CTI and sacrificial charge (bottom)}
\label{fig:phvsrow}
\end{figure}

Figure~\ref{fig:phvsrow} is a scatter plot of the pulseheight of each X-ray event versus its row number for the three cases of no correction, standard CTI correction and CTI plus sacrificial charge correction.  Each of the near horizontal lines represents a spectral feature in the ACIS calibration source.  In the absence of radiation damage, the spectral features would appear as horizontal lines with no change in the line width with increasing row number.  For the uncorrected data, the pulseheight of each spectral line drops with increasing transfer distance.  The spectral resolution is also a function of position and degrades with increasing transfer distance.  Near the framestore, at low row numbers, the performance of the device is essentially the same as before the radiation damage.  Standard CTI correction removes the position dependence of pulseheight and improves the spectral resolution.  The addition of the sacrificial charge correction further improves the spectral resolution which can best be seen by the better separation of the Mn and Ti K-$\alpha$ and K-$\beta$ lines.

\begin{figure}
\vspace{3in}
\includegraphics{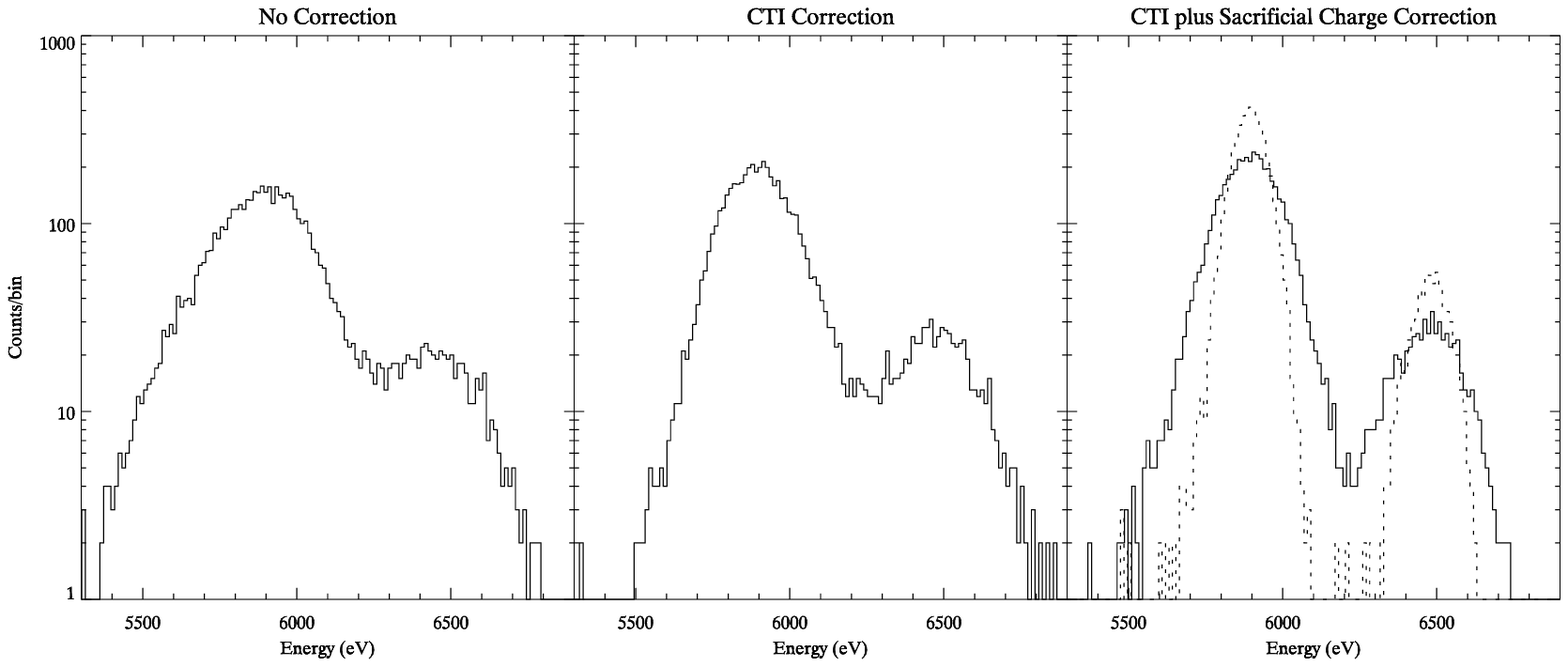}
\caption{Pulseheight spectra around 6~keV for data above row~800 that is uncorrected (left), has been corrected for CTI (center), and has been corrected for CTI and sacrificial charge (right).  The dotted line shows the original undamaged performance.}
\label{fig:spec}
\end{figure}

Figure~\ref{fig:spec} shows the pulseheight spectra around 6~keV for data above row 800.  Also shown is the original, undamaged performance of the CCD.  The addition of the sacrificial charge correction removes the high energy shoulder of the response and clearly improves the separation of the Mn-K$\alpha$ and K$\beta$ lines over the CTI correction alone.

\begin{figure}
\vspace{4.0in}
\includegraphics{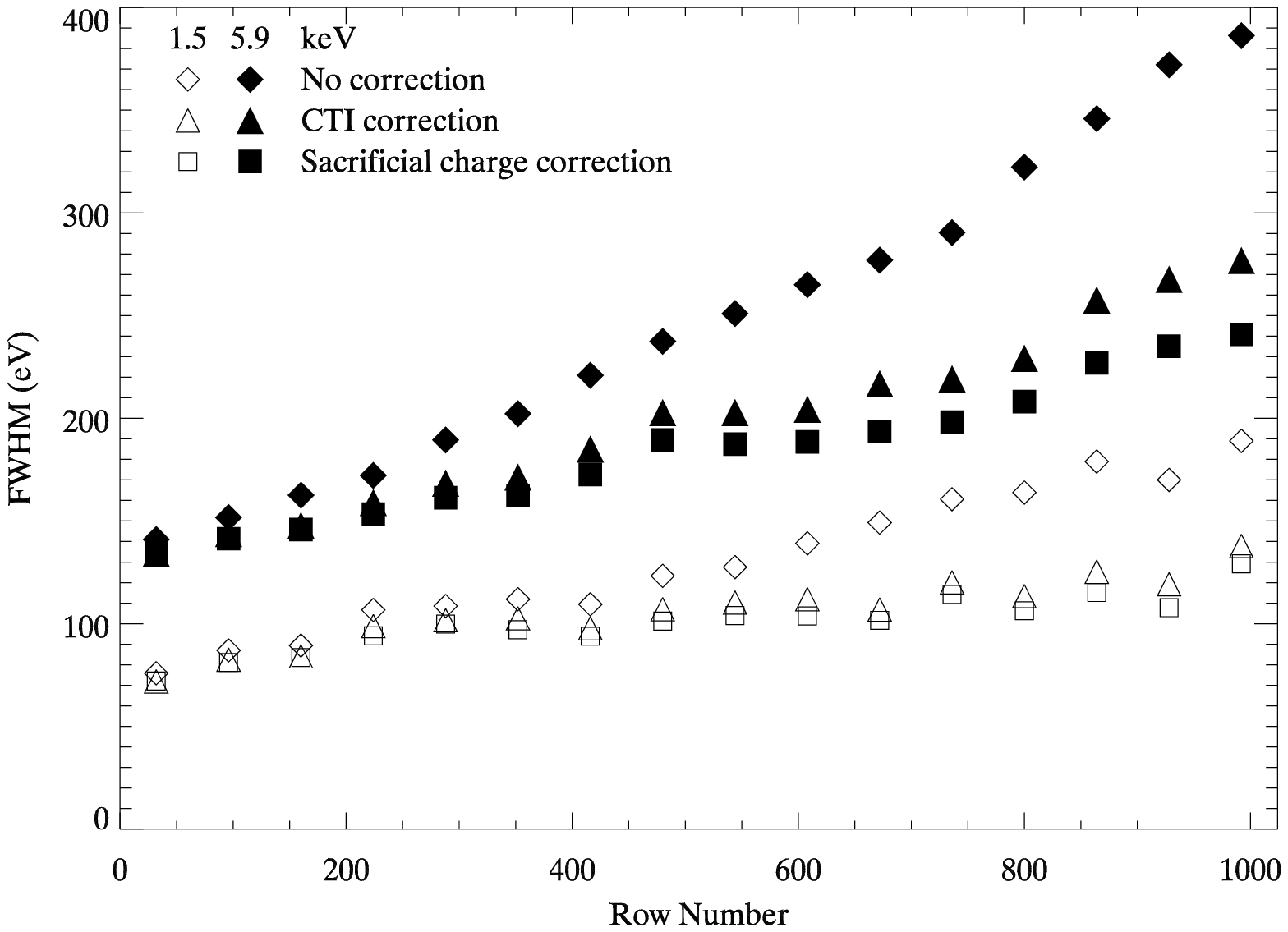}
\caption{FWHM of the spectral lines at 5.9 keV and 1.5 keV as a function of row number for uncorrected, CTI corrected and sacrificial charge corrected data.}
\label{fig:fitrow}
\end{figure}

Figure~\ref{fig:fitrow} shows the FWHM of the spectral lines at 5.9 and 1.5~keV as a function of row number.  At low row numbers, the FWHM approaches the undamaged value.  Applying the sacrificial charge correction improves the FWHM by up to 13\% at 5.9~keV and up to 10\% at 1.5~keV over the CTI correction alone.  At the highest row bin, the FWHM at 5.9~keV improves from 388~$\pm$~6~eV with no corrections, to 277~$\pm$~3~eV with CTI correction alone, and to 243~$\pm$~3~eV with CTI and sacrificial charge correction.  The improvement is smaller at 1.5~keV, even though one might expect that sacrificial charge would become more important at lower energies.

\section{DISCUSSION}
\label{sect:disc}

\begin{figure}
\vspace{4.0in}
\includegraphics{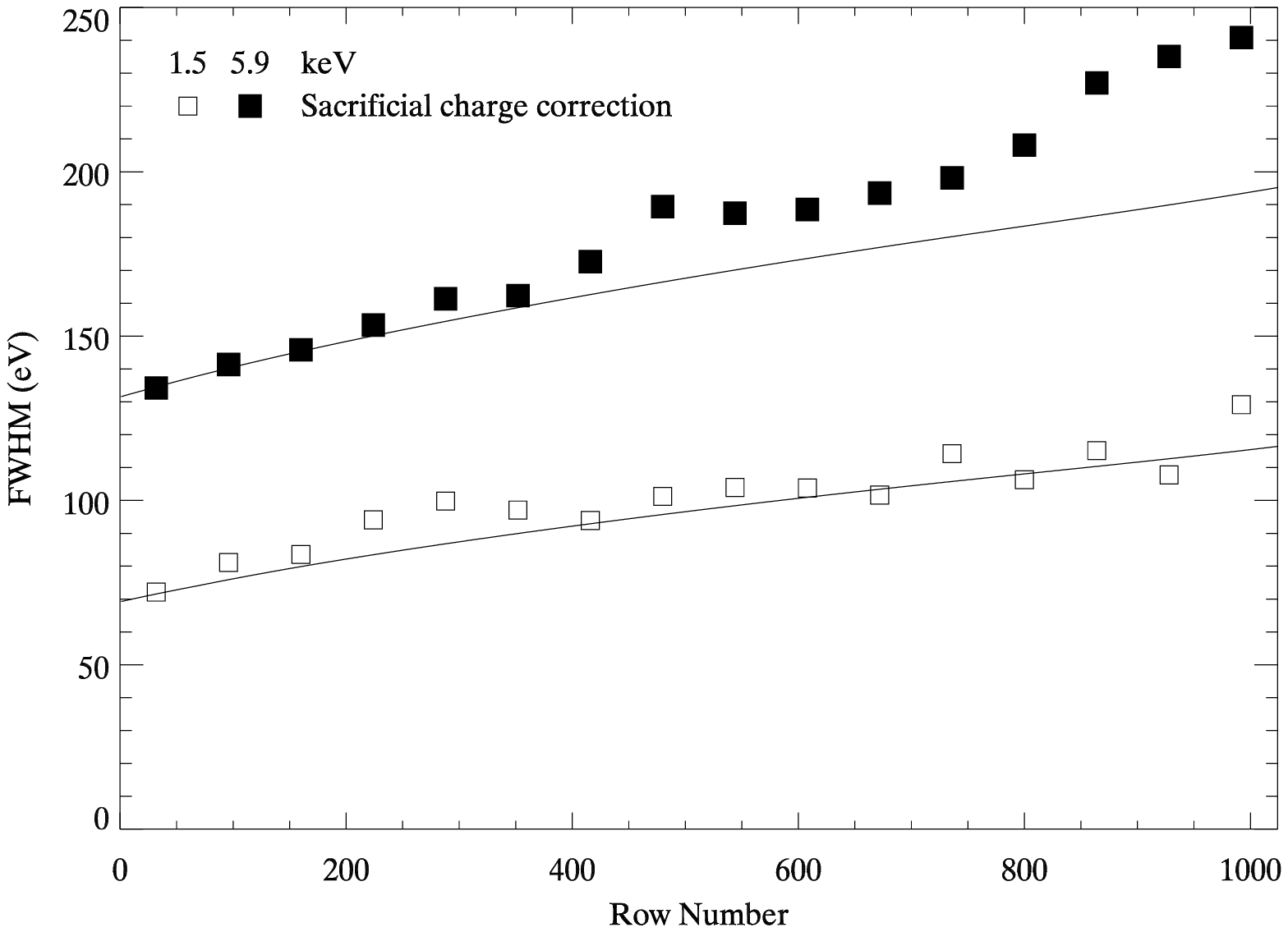}
\caption{FWHM of the spectral lines at 5.9 and 1.5~keV as a function of row number for sacrificial charge corrected data.  The solid line is an estimate of the theoretical limit on performance.}
\label{fig:ideal}
\end{figure}

While our preliminary CTI plus sacrificial charge correction algorithm does offer improved performance, we believe that further improvement is possible.  Figure~\ref{fig:ideal} compares the FWHM at 5.9 and 1.5~keV versus row number resulting from our sacrificial charge correction with an estimated theoretical limit on performance.  This limit was determined by assuming that the fluctuations in charge loss and in sacrificial charge are Poissonian and includes an estimate of the additional noise from split events.  At 1.5~keV the performance has reached this theoretical limit, while at 5.9~keV the sacrificial charge corrected data are still substantially worse than the limit.  

One source of excess noise in the sacrificial charge correction is the presence of cosmic ray events which interact in the framestore region of the CCD.  These events appear in raw frames and event lists as sacrificial charge and have been included in the calculation of $Z_X$, but do not actually influence the pulseheight of X-ray events.  If a framestore event is present, particularly if it appears to be the closest precursor charge to the X-ray event, the value of $Z_X$ will be overestimated and the charge loss underestimated.  These X-ray events will be poorly corrected and should broaden the spectral resolution.  While it is possible to run the CCD such that only framestore events are read out to determine general population characteristics, it is not possible to know which precursors of a given X-ray are sacrificial charge and which are not.

Another possible source of performance degradation that is ignored in the current treatment is the influence of sacrificial charge from previous integration frames of the CCD.  The typical exposure time for an ACIS CCD frame is 3.2~sec, while the longest measured emission time constant on ACIS is estimated at 4 - 6~sec.  While events with sacrificial charge in the same frame should be relatively unaffected, some events with $Z_X$ = 0 may have their charge loss overestimated and should again broaden the spectral resolution.

\begin{figure}
\vspace{4.0in}
\includegraphics{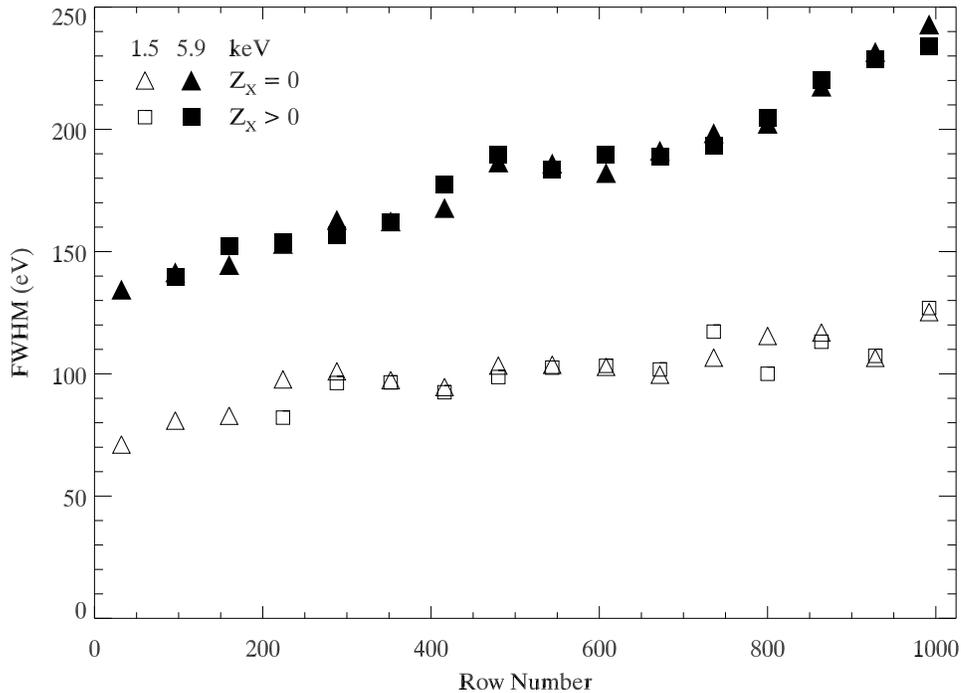}
\caption{FWHM of the spectral lines at 5.9 and 1.5~keV as a function of row number for events corrected for CTI and sacrificial charge for the case of $Z_X > 0$, events with sacrificial charge, and of $Z_X = 0$, events with no sacrificial charge.}
\label{fig:znoz}
\end{figure}

To check the efficacy of further work on the sacrificial charge algorithm alone, Figure~\ref{fig:znoz} is a comparison of the FWHM versus row number for events with $Z_X > 0$ (sacrificial charge present) to events with $Z_X = 0$ (no sacrificial charge) after CTI plus sacrificial charge correction.  The results are not significantly different, indicating that the events with sacrificial charge have been corrected as well as events that do not have sacrificial charge within the statistical errors of the data. Since the results in Figure~\ref{fig:ideal} argue that more improvement is possible, at least at higher energies, we must examine the assumptions in both the CTI and sacrificial charge correction.

The primary difference between the low and high energy events is the importance of events split into multiple pixels; at 1.5~keV 79\% of the events are single pixel, 19\% are doubly split and 2\% are multiply split, while at 5.9~keV 47\% are singly split, 33\% are doubly split and 20\% are multiply split.  The additional spectral resolution degradation at 5.9~keV over 1.5~keV may be a result of incorrect treatment of split events.  There are at least two means by which split events could be poorly corrected.  The treatment of self-shielding within the event island assumes that the emission time constants are much longer than the event transfer time (40~$\mu$sec), however there are a small number of traps with a time constants close to the transfer time (2\% with 60~$\mu$sec).  This should be a small effect, since the fraction of relevant traps is so small.  The energy dependence of charge loss, which is currently parameterized as a power law, has not yet been confirmed at the lowest energies ($<$~670~eV).  Split events will often include pixels with small amounts of charge, so an error in the energy dependence could produce a additional noise in the corrected pulseheights.  This possibility is currently under investigation.

If a sacrificial charge correction scheme does provide significant performance improvement, its implementation will require changes to the flight software to telemeter additional information about sacrificial charge history with each event without making a significant impact on telemetry bandwidth.  Such a patch is under development by the ACIS team.  The strategies currently under study include encoding precursor information into bits normally used to telemeter corner pixel (in 3x3 faint mode) or outer pixels (in 5x5 very faint mode) pulseheight.

\acknowledgments

We would like to thank our colleagues at MIT and the CXC especially Peter Ford and Steve Kissel, also Leisa Townsley and her coworkers at Penn State who developed ACIS CTI correction techniques upon which our own are based.  This work was supported by NASA contracts NAS8-37716 and NAS8-38252.

\bibliography{report}   
\bibliographystyle{spiebib}   

\end{document}